\documentclass[journal=jacsat,manuscript=article,layout=twocolumn]{achemso}
\usepackage[version=3]{mhchem} 
\usepackage[T1]{fontenc}       
\usepackage{color,soul} 
\usepackage{textcomp}
\usepackage{listings}
\definecolor{mygreen}{RGB}{28,172,0} 
\definecolor{mylilas}{RGB}{170,55,241}

\newcommand*{\sometext}{Single-mode optical nanofibres are a central component of a broad range of applications and emerging technologies. Their fabrication has been extensively studied over the past decade, but imaging of the final sub-micrometre products has been restricted to destructive or low-precision techniques.  
Here we demonstrate an optical scattering-based scanning method that uses a probe nanofibre to locally scatter the evanescent field of a sample nanofibre. The method does not damage the sample nanofibre and is easily implemented only using the same equipment as in a standard fibre puller setup. We demonstrate sub-nanometre radial resolution at video rates (0.7~nm in 10~ms) on single mode nanofibres, allowing for a complete high-precision profile to be obtained within minutes of fabrication. 
The method thus enables non-destructive, fast and precise characterisation of optical nanofibers, with applications ranging from optical sensors and cold atom traps to non-linear optics.  
}

\let\oldmaketitle\maketitle
\let\maketitle\relax
\author{Lars S. Madsen}
\email{m.lars@uq.edu.au}
\affiliation[UQ]
{Centre for Engineered Quantum Systems, School of Mathematics and Physics, The University of Queensland, St. Lucia, Brisbane, Queensland 4072, Australia}
\author{Christopher Baker}
\affiliation[UQ]
{Centre for Engineered Quantum Systems, School of Mathematics and Physics, The University of Queensland, St. Lucia, Brisbane, Queensland 4072, Australia}
\author{Halina Rubinsztein-Dunlop}
\affiliation[UQ]
{Centre for Engineered Quantum Systems, School of Mathematics and Physics, The University of Queensland, St. Lucia, Brisbane, Queensland 4072, Australia}
\author{Warwick P. Bowen}
\affiliation[UQ]
{Centre for Engineered Quantum Systems, School of Mathematics and Physics, The University of Queensland, St. Lucia, Brisbane, Queensland 4072, Australia}
\title[title]
  {Non-destructive profilometry of optical nanofibres}

\abbreviations{SEM,ONF}
\begin{document}
%
%
%
%
%
\twocolumn[
\begin{@twocolumnfalse}
\oldmaketitle
\begin{abstract}
\sometext
\end{abstract}
\end{@twocolumnfalse}
]
Optical nanofibres (ONFs) offer a wide variety of applications, ranging from optical sensors to nonlinear components and couplers to plasmonic and optomechanical systems [see \cite{tong2012optical} and references herein]. They also form the basis for emerging experimental platforms within cold atom physics \cite{vetsch2010optical,goban2012demonstration,kato2015strong,beguin2014generation,gouraud2015demonstration,lee2015inhomogeneous} as well as in nano- and bio-particle detection \cite{yu2014single,swaim2013tapered,nico2016bio}. The workhorse in these experiments is the intense evanescent optical field extending out of the optical nanofiber. For instance, these fields have allowed the generation of optical potentials for trapping cold atoms with a single atom optical depth as high as a few percent \cite{goban2012demonstration}; while in nano- and bio-imaging, trapping and detection of 5~nm silica particles and 4~nm bio-molecules in solution has been achieved \cite{nico2016bio}. In such applications, a uniform evanescent field is often required in order to ensure constant interaction strength over the length of the ONF.

Several techniques are available to fabricate high transmission ONFs, based on heating and pulling or etching a standard step index optical fibre \cite{hoffman2014ultrahigh,ding2010ultralow,nagai2014ultra,chenari2016adiabatic,tong2003subwavelength}. For the heat-based methods, the theory describing the average resulting fibre shape is well understood \cite{birks1992shape} and likewise is the relationship between the shape and the transmission \cite{love1991tapered}. Experimentally ONFs with transmission greater than $99\%$ have been demonstrated \cite{hoffman2014ultrahigh,ding2010ultralow,nagai2014ultra}. Often however, fluctuations in the fabrication process are responsible for deviations from the desired shape. Fluctuations occurring early in the pulling process affect the optical transmission and can thus be identified directly during fabrication. On the other hand, fluctuations occurring later in the process can lead to an inhomogeneous evanescent field, without appreciably affecting the overall transmission levels, making them difficult to identify.

The general approach to characterise the evanescent field of an ONF has been to image the diameter of the ONF along its length. Imaging of the ONFs with standard microscopy techniques is difficult since the ONF diameter is comparable to the diffraction limit. Scanning electron microscopy (SEM) can give very high resolution in both the radial and the axial directions, but it is challenging to use the technique without damaging the ONF. SEM images have been used to determine the statistical reproducibility of the fabrication methods and generally show that ONFs are structured with variations in diameter of several percent \cite{brambilla2004ultra,tong2003subwavelength,garcia2011optical,goban2012demonstration,kato2015strong,gusachenko2015optical}. The SEM images are generally taken sparsely along the ONFs as continuous imaging of a millimetre long ONFs would require several hundred images per millimetre. Due to the limitations of SEM imaging a variety of alternative methods have been employed to image ONFs \cite{wiedemann2010measurement, holleis2014experimental, coillet2010near, hoffman2015rayleigh,gusachenko2015optical, sumetsky2006,sumetsky2010, keloth2015}. Early work by \citeauthor{sumetsky2006} demonstrated that the ONF radius could be determined with nanometre resolution by scanning a partly stripped fibre through its evanescent field and measuring its the transmission. \cite{sumetsky2006} The method was limited by drag between the probe fibre (125 $\mu$m radius) and the sub-wavelength ONF and relied upon SEM images for calibration, hindering widespread application \cite{sumetsky2010}. Later approaches to measure ONFs include stress strain analysis, nonlinear higher frequency generation, scanning nearfield optical microscopy, use of microcavities and analysis of multimode interference patterns\cite{wiedemann2010measurement, holleis2014experimental, coillet2010near, hoffman2015rayleigh,gusachenko2015optical, sumetsky2010, keloth2015}.

In this paper we return to the idea of measuring transmission with a scatterer placed in the evanescent field of a sample ONF. We overcome the problem of drag by using microfibres as probe fibres. Further, we find that the multimode behavior in the tapered section provides a mean for absolute calibration of the radius eliminating the need for SEM images and thus making the method fully nondestructive. The technique is easily incorporated in standard heat-and-pull setups and can be used independently of the ultimate application. In practice, images of millimetre long ONFs can be obtained within a few minutes of fabrication, allowing for immediate quality control. We obtain a radial resolution of 0.7~nm and an axial resolution  of around 1 micron. We observe that even high transmission ONFs can have fluctuations of around $5\%$ in their radius on length scales of tens of micrometres, adversely affecting the uniformity of the evanescent field.


\begin{figure}[htb]
\includegraphics[width=0.5\textwidth,keepaspectratio]{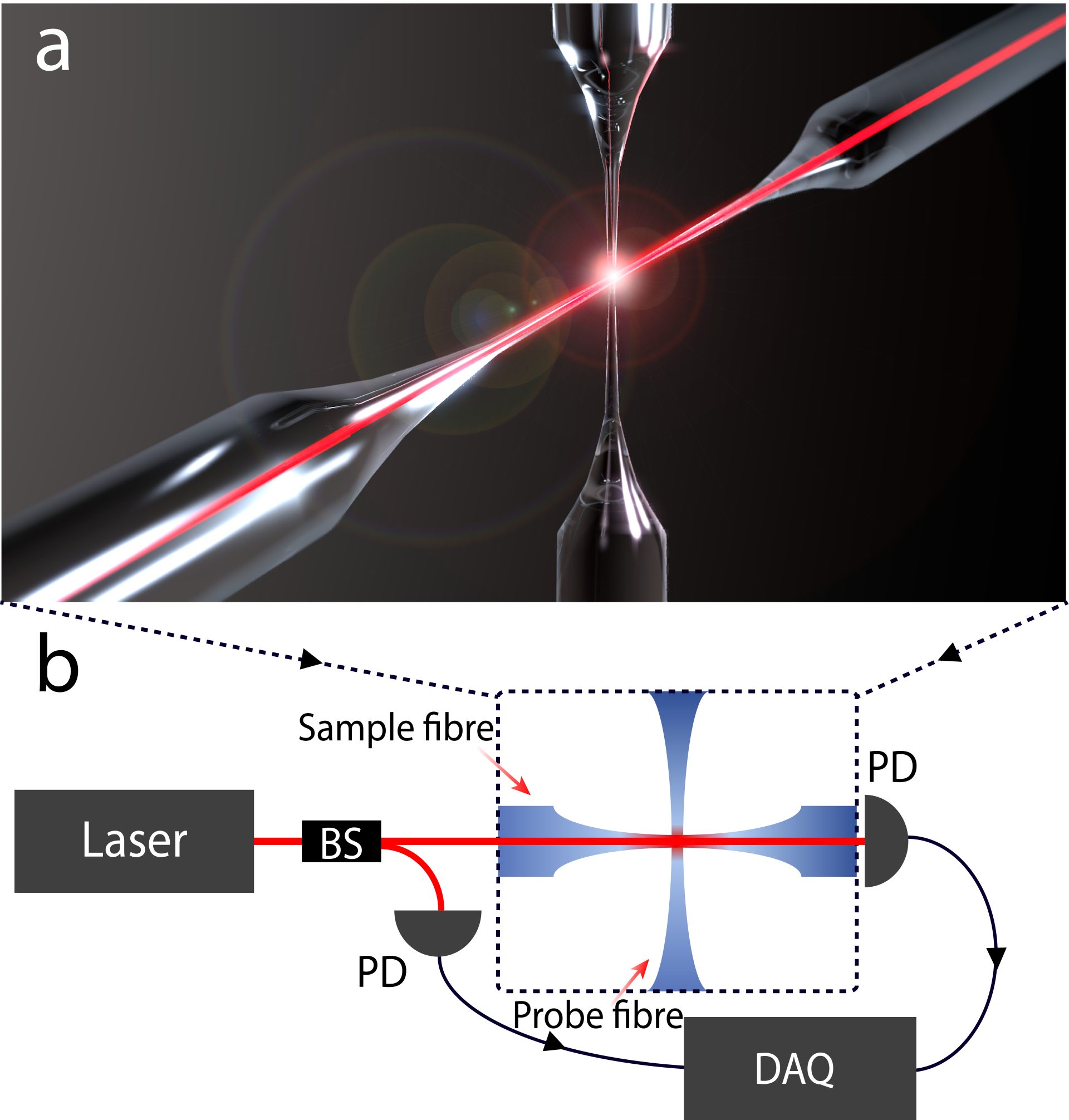}
\caption{ Illustration of the experimental setup. a) Artistic illustration. b) Experimental setup. Light from a 780~nm laser is split with a $50/50$ beamsplitter (BS). Half of the light is measured for intensity stabilisation with a photodetector (PD) and half is coupled into the sample fibre. A second probe fibre is placed in contact with the sample and stepped horizontally along the waist while the transmission is monitored with a second photodetector (PD). The signals from the two photodetectors are digitized with a data acquisition system (DAQ).
}
\label{fig1}
\end{figure}


\begin{figure*}[t]
\includegraphics[width=1\textwidth,keepaspectratio]{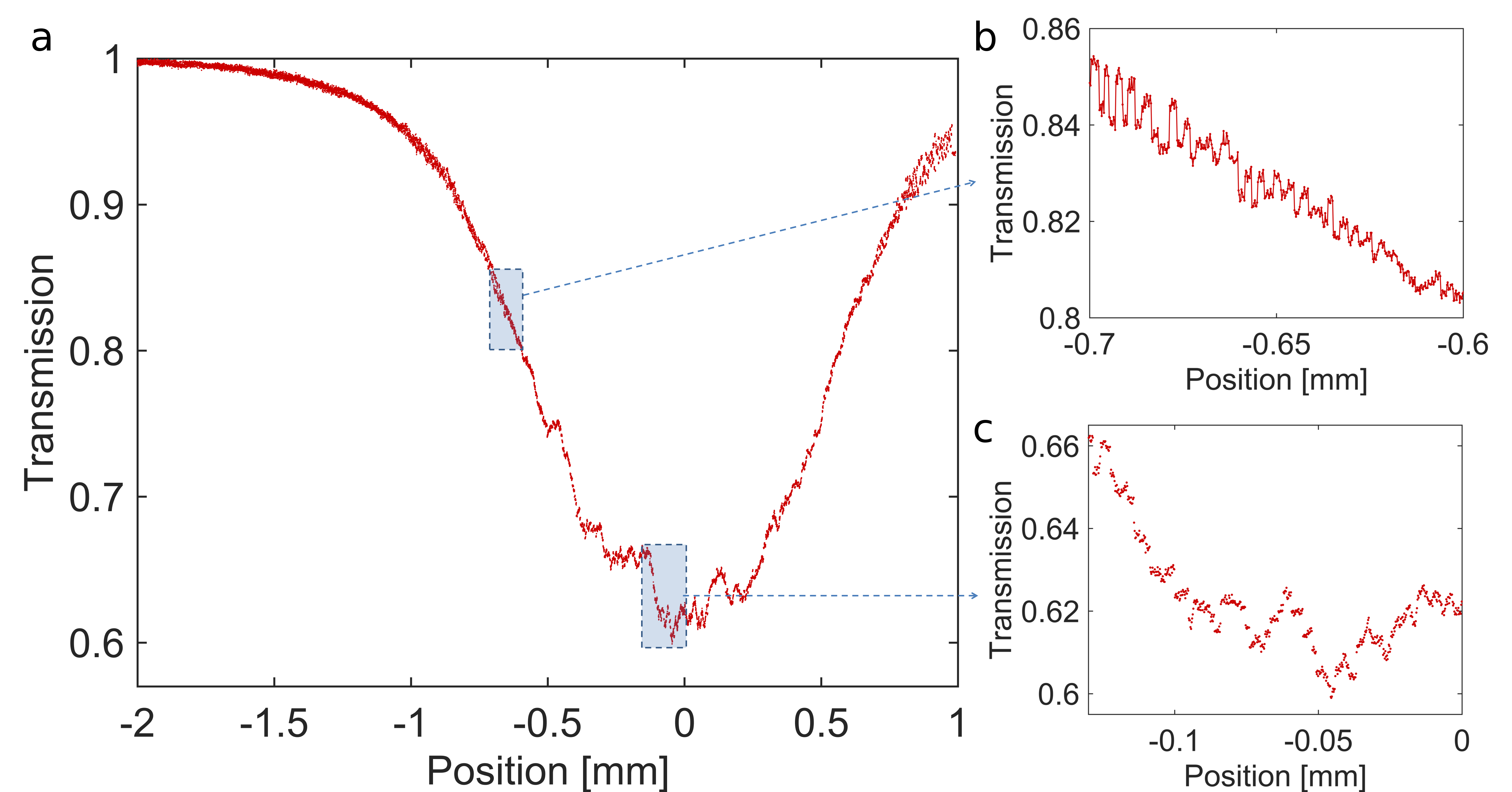}
\caption{Transmission data. Normalised transmission measured as function of the probe fibre position along the sample "fibre 1". a) Full profile of the ONF region. Each of the data points is an average of 10 ms measurements with 100 kHz sampling rate.  b) Zoom in on the region where the fibre becomes single mode. Measurement points have been connected as a guide to the eye. c) Zoom in on the large fluctuation near the waist. 
}
\label{fig2}
\end{figure*}

Our approach to obtain a profile of a nanofibre, referred to as the \textit{sample fibre}, is to inject light into it and step a second tapered \textit{probe fibre} along it with the two in contact while measuring the transmission, as shown in Figure \ref{fig1}. When the probe fibre is in the thin region of the sample fibre, where the evanescent field extends out into the surrounding air, a fraction of this light is scattered. By measuring the resulting reduction in transmission an estimate of the evanescent field intensity at the probe fibre position is obtained.

Figure \ref{fig2} a) shows an example of a transmission measurement when stepping the probe fibre along a sample fibre, with a 10 ms measurement time at each step and 0.2~$\mu$m step size (See Supporting Information for details). As expected, we see that far from the ONF region the probe has no effect on transmission. As the probe approaches the nanofibre section, the transmission drops to a minimum near $60\%$ of its maximal value. On the slopes of the tapered region a periodic modulation is observed that dies out when the transmission decreases below $81\%$, as shown in the zoomed portion of the plot in Figure \ref{fig2} b). We attribute this modulation to interference in the multimode section of the ONF caused by the difference in propagation constants between the fundamental and first higher order guided modes, as also observed by \citeauthor{hoffman2015rayleigh} \cite{hoffman2015rayleigh} At the point where the oscillation dies out and the fibre becomes single mode theory predicts the silica fibre radius to be $0.36 \lambda$ \cite{le2004field}, providing a fixed point for absolute calibration between the measured transmission and the nanofibre radius. Unexpectedly the transmission in the central nanofibre region is not smooth. As can be seen in figure \ref{fig2} c), the transmission is highly structured with a variation up to $ 7\%$ over 30~$\mu$m along the ONF and finer structures giving rise to changes in transmission of around $1\%$. For applications requiring a uniform evanescent field such as a cold atom trap this particular ONF would be non-ideal despite no obvious signs of defects during fabrication and $98\%$ transmission. 

\begin{figure*}[htb]
\includegraphics[width=1\textwidth,keepaspectratio]{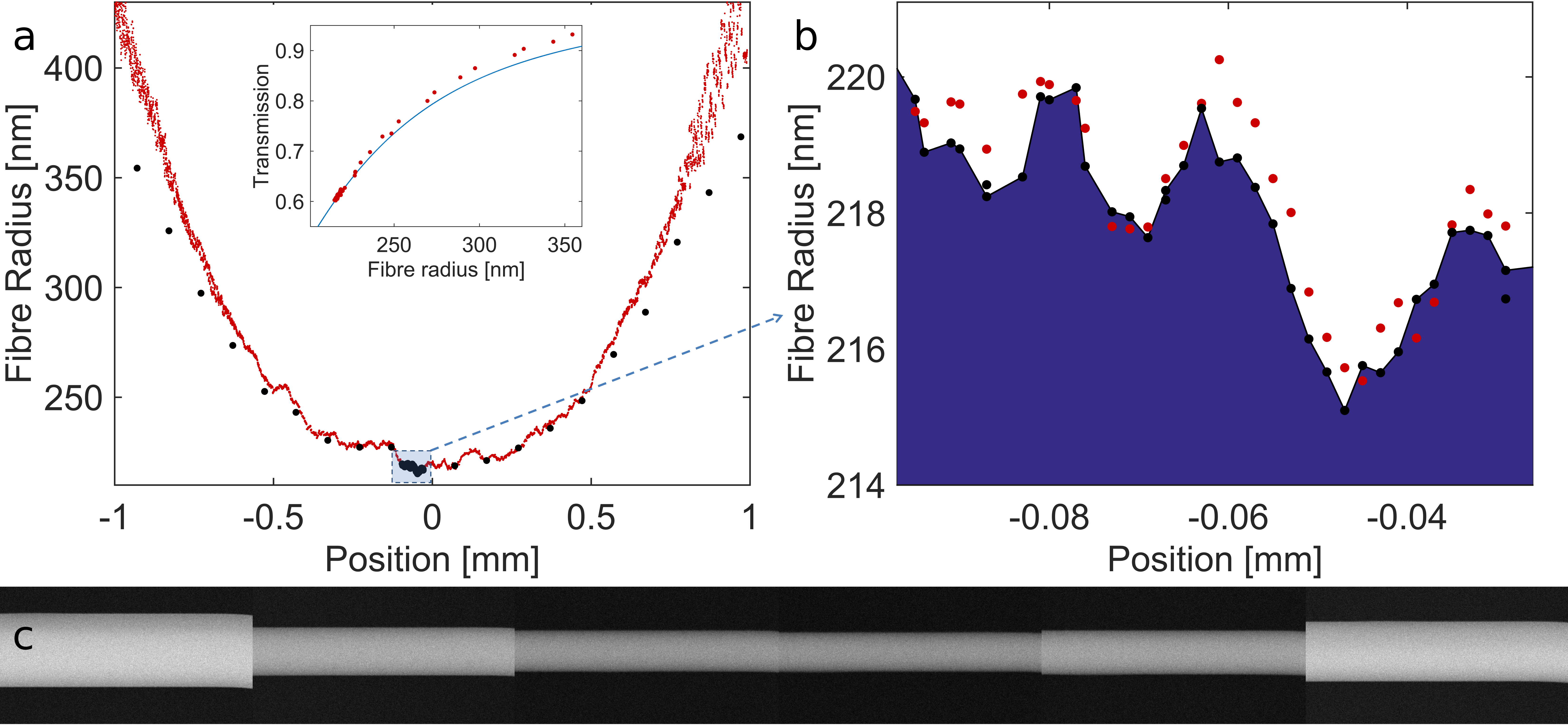}
\caption{Comparison with SEM measurements. a) The transmission data in Figure \ref{fig2} is used to derive the fibre radius using Eqn 1 with $\eta=1.00$; results shown with the red points. The black points mark the SEM data and the grey shading under the points is a guide to the eye. Each point is the average radius on one image. Inset: Red points show the average over 3~$\mu$m bins of transmission measurements as function of the SEM measured thickness. The blue line is modelled with $\eta=1.00$. b) Zoom in on the region from a) near the waist with dense SEM imaging. The scan data is averaged in 3~$\mu$m bins centred at the SEM positions. c) 6 juxtaposed examples of the SEM images used to measure the sample fibre radius, each image covers 3 $\mu$m  of the fibre and is averaged giving one black point. 
}
\label{fig3}
\end{figure*}

To quantify the relation between the ONF radius and the transmission measurements we develop a simple phenomenological model. ONFs can be modelled as step index fibres where the core is glass and the cladding is air\cite{snyder2012optical,le2004field,vetsch2010thesis}. This model predicts the fraction of optical power that is inside the ONF $P_{in}(r)$ and the part which is contained in the evanescent field $P_{out}(r)$ as function of the ONF radius $r$ (See Supporting Information for details). To model the reduction in transmission $T(r)$ caused by placing a probe fibre onto it, we assume that for a given probe a fixed fraction $\eta$ of the power in the evanescent field is scattered from the guided mode so that: 

\begin{equation}
T(r)=\frac{P_{in}(r)+ (1-\eta)P_{out}(r)}{P_{total}(r)}, \label{eqn1}
\end{equation}
where $P_{total}(r)=P_{in}(r)+P_{out}(r)$ is the total power. The fraction of power outside the ONF at the point where the fibre becomes single mode is 19\% so $\eta =(1-T_{sm})/0.19$ can be determined experimentally in situ for a given probe fibre by determining the transmission ($T_{sm}$) at that point, as shown in Figure \ref{fig2} b). We note that due to coupling between the inside and outside fields $\eta$ can exceed unity, as discussed in the Supporting Information.

To confirm that the observed structures in the evanescent field are indeed caused by fluctuations in the radius of the ONF we image the sample fibre with a scanning electron microscope (SEM). Evenly spaced images each covering  3~$\mu$m of the ONF were taken every 100~$\mu$m along the tapered region of the fibre, as well as closely spaced images covering the fibre continuously for 65~$\mu$m near the fibre waist. The width of the ONF on the SEM images is digitally recognised with a systematic uncertainty of $\pm10$ nm, (as discussed in the Supporting Information). The data of each SEM image is averaged to obtain one measurement of radius and the optical scan data of Figure \ref{fig2} is converted into a profile of the sample fibre radius using Eq. (\ref{eqn1}). Figure \ref{fig3} a) shows the SEM data overlayed with the sample radius derived from the optical scan method. Taking the mean of the optical scan data in bins corresponding to the position and size of the SEM images the two sets of data can be statistically compared, giving a correlation coefficient exceeding 0.99. While the model fits for the single mode region of the fibre it starts to deviate in the multimode region. We expect the deviation to be caused by the simplicity of our model which assumes the same fraction of the evanescent field to be scattered in both the single mode and the multimode region. A full model is beyond the scope of this paper. Figure \ref{fig3} b) shows the region of the nanofibre which has been densely imaged with SEM together with the averaged bins of optical scan data. The optical scan data and the SEM data have an average offset of around 1~nm. The correlation coefficient in this region is 0.94 making it apparent that the small fluctuations in transmission are primarily caused by fluctuations in the radius of the ONF sample.

\begin{figure}[htb!]
\includegraphics[width=0.42\textwidth,keepaspectratio]{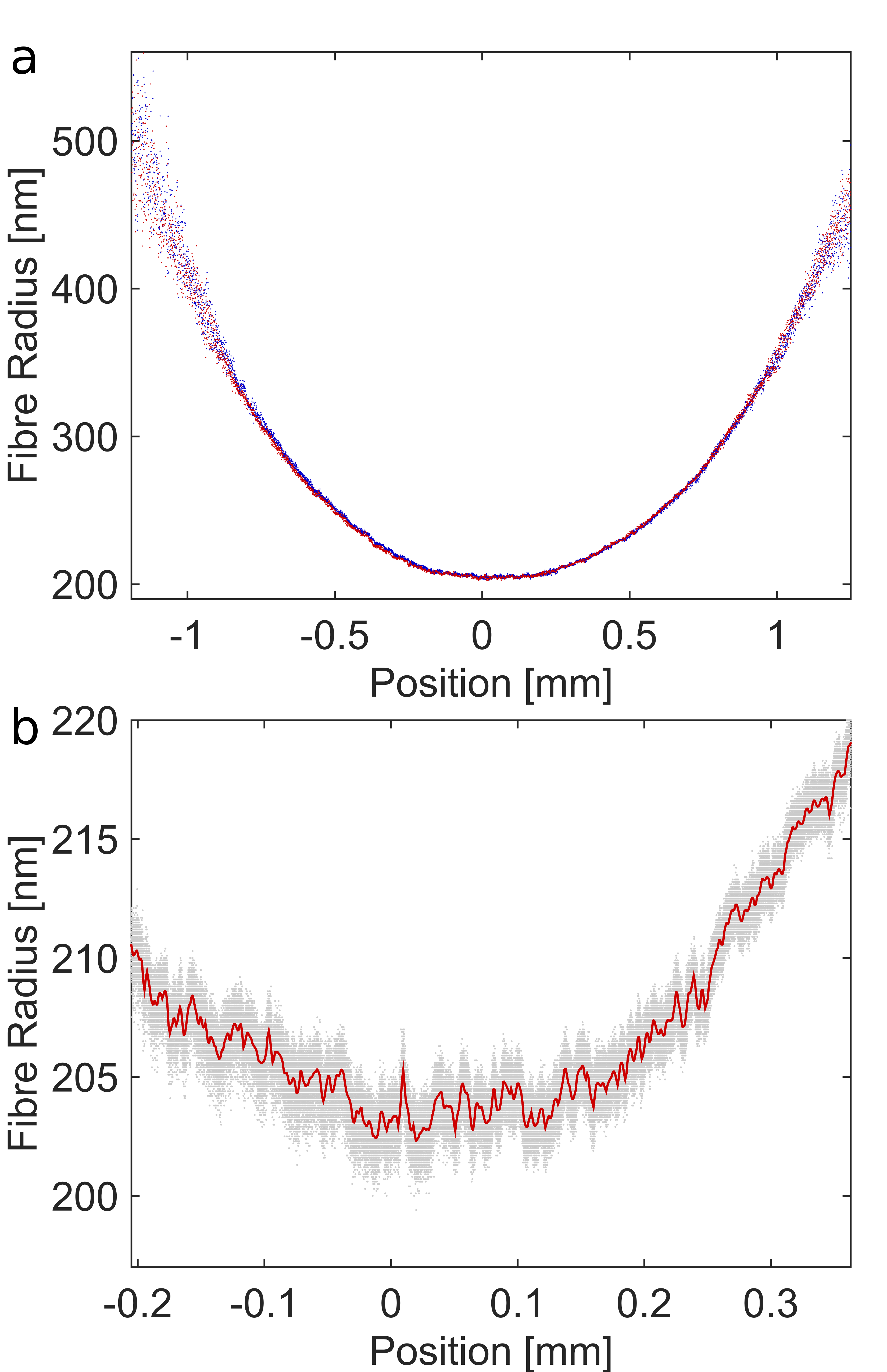}
\caption{ Data for statistical analysis of the technique. a) The radius derived from 4 full scans of the second sample with $\eta=0.82$, left to right scans (red points) have been compensated for 4~$\mu$m of hysteresis relative to right to left scans (blue points). b) Data for the 302 scans near the waist of the fibre marked with grey points. The data has been compensated for hysteresis and scaled to fix the average of the 100~$\mu$m right most data points. The mean of all the scans is shown with the red line. 
}
\label{fig4}
\end{figure}

To quantify the resolution and reproducibility of the technique we profile a second ONF in greater detail. First we profile the full tapered section and evaluate the radius, shown on Fig. \ref{fig4} a). This second nanofibre is much more uniform than the first, even though both were fabricated with very similar procedure and both exhibit $98\%$ transmission (See Supporting Information for details). This provides an example of why characterisation beyond the standard transmission measurements done during fabrication is essential to applications that rely on uniformity of the evanescent field. We scan both left to right and right to left and observe a 4~$\mu$m hysteresis, which we attribute to the drag between the sample and the probe. This hysteresis has been compensated for in both Fig \ref{fig4} a) and b). Second we profile 0.6 mm of the central waist section 302 times in succession. During the measurement time of 4 hours the transmission drops a total of $2.1\%$ which is comparable to what we would expect without profiling the fibre, emphasising the non-destructive nature of the method. To correct for this systematic drift we scale each trace to fix the transmission to the average of the right most 100~$\mu$m of data. As can be seen in Figure \ref{fig4} b) the measured radius is highly reproducible. The structures along the ONF axis are smooth on a length scale of around 1~$\mu$m setting an upper bound on the axial resolution. The standard deviation of the transmission is 0.003 of an average transmission of 0.631, corresponding to radial resolution in a single 10~ms measurement of 0.7~nm.

To test the applicability of the technique when changing the probe thickness the second ONF is imaged with different thicknesses of the probe fibre giving transmissions at the waist between $80\%$ and $10\%$ (See Supporting Information for details). Thicker probes are observed to provide significantly higher radial resolution, but the increased drag introduces irregular hysteresis effects. Deriving the radius of the ONF using Eq.  \ref{eqn1} for different probe thicknesses gives consistent results for the radius in the waist region to within $5\%$. 


Finally we investigate the profile dependence on the input polarisation. The polarisation is observed to change the relative transmission by up to 10\% at the waist  (See Supporting Information). Maintaining a stable input polarisation during the experiment is therefore essential. We find that near identical ONF profiles are obtained, independent of input polarisation. This shows that our method is robust to the input polarisation and that the input polarisation remains stable in the experiment. 
Furthermore the method allows characterisation of the birefringence of the sample. When profiling the ONF with several polarisation inputs the birefringence in the ONF would modify the observed profiles (as discussed in the Supporting Information). The experiment thus shows that the characterised ONF has negligible birefringence.

The axial and radial resolution obtained here is sufficient to characterise the evanescent field uniformity of current fabrication techniques. In future applications, if improved resolution is required, the current limitations are mainly technical set by the laser noise and polarisation fluctuations in the fibre. Further intensity stabilisation, longer measurement time and polarisation sensitive detection could bring the sensitivity deeper into the sub-nanometre scale. A different approach to improving the resolution is to experiment with the probe. As shown thicker probes can give improved resolution, but at a cost of increased drag effects. Coating a thin probe fibre with a high contrast material could provide higher scattering without causing increased drag.  

The size of the structures in the nanofibre surfaces observed here is consistent with previous SEM studies of fibre surface quality \cite{brambilla2004ultra,tong2003subwavelength,garcia2011optical,goban2012demonstration,kato2015strong}. The full profiles of ONFs show that these structures can cover the entire ONFs rather than being isolated incidents. Our technique facilitates future experiments to scrutinise the cause of these structures. 

In summary, we demonstrate a method for quick, precise and non-destructive characterisation of optical nanofibres. With this method a complete width profile of an optical nanofibre can be measured with sub-nanometre resolution within minutes of fabrication and independent of the fibres final application. The availability of easy and fast characterisation will allow highly uniform ONFs to be selected for atom trapping and nonlinear optics, bio sensors to be calibrated, and fabrication methods to be optimised.

\subsection{ASSOCIATED CONTENT}

\subsection{AUTHOR INFORMATION}
\subsubsection{Corresponding Author}
*E-mail: m.lars@uq.edu.au
\subsubsection{Present Addresses}
Building 6, School of Mathematics and Physics, The University
of Queensland,  St. Lucia, Brisbane, Queensland 4072, Queensland, Australia.

\subsubsection{Notes}
The authors declare no competing financial interest.

\subsection{ACKNOWLEDGMENTS}
This work was funded by the Australian Research Council through the Centre of Excellence for Engineered Quantum
Systems (EQuS, CE110001013), and by  the  Air  Force  Office  of  Scientific  Research  and  the Asian  Office  of  Aerospace  Research  and  Development. W.P.B. is supported by the Australian Research Council Future  Fellowship  FT140100650.

\bibliography{nearfield}

\end{document}


%
%
%
%
%
\oldmaketitle
\section{Methods}
\begin{figure}[htb]
\includegraphics[width=0.5\textwidth,keepaspectratio]{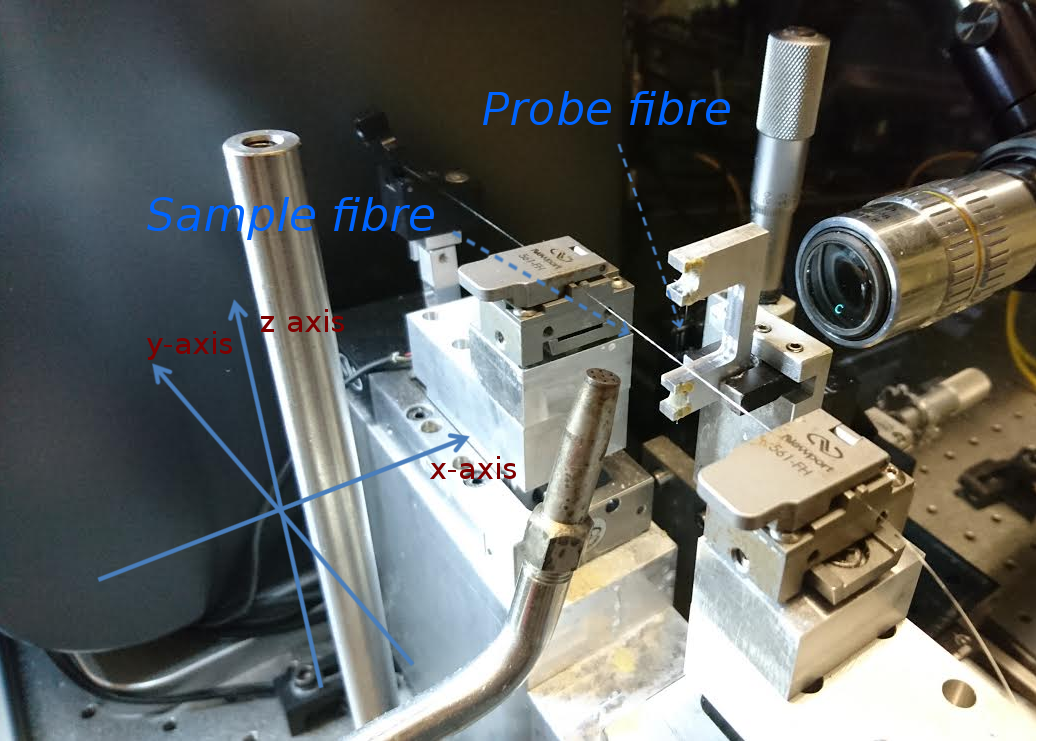}
\caption{Experimental configuration of the probe and sample fibres. The sample fibre here is not tapered to make it easier to see.  
}
\label{fig1sup}
\end{figure}

The imaging method is implemented in our fibre puller setup without requiring any major additional equipment. The two computer controlled micrometre stages (Newport 561-FH on MFA CC via a ESP 300) which are used to hold and stretch the fibre while tapering are here used to move the sample fibre relative to the probe fibre. The probe fibre is a tapered fibre, glued to an orthogonal fibre clamp that can be brought into contact with the sample fibre with a manual xz-micro-meter stage, see Figure \ref{fig1sup}. 780 nm light from a Velocity 6312 diode laser is split in two parts with a fibre beamsplitter. One part is measured directly with a Thorlabs PDA10CS-EC photodetector for intensity stabilisation. The other part is sent through the sample fibre and then measured on a identical detector. To minimize back reflection all fibre connections are angle polished (FC/APC) except the output of the sample which is just cleaved. The electronic signals from the detectors are digitalised with a DAQ (NI PCI 6221). The probe is positioned to scatter ~40$\%$ of the total light when on the waist of the sample. The scattering is adjusted by changing the thickness of the probe fibre at its contact point with the sample fibre through the z-micrometer stage. The 40$\%$ is a compromise between high signal to noise ratio and low hysteresis as discussed later. Both sample and probe need to be lightly tensioned to minimise hysteresis. 

In the scan data presented for fibre 1 we use a step size of 0.2 um and 10 ms measurement time. We include a 20 ms pause between moving the probe fibre and measuring the transmission through the sample fibre, in order to allow any vibration caused by the motion of the probe to die out. Each of the transmission data points in figures 2-4a is the average of 1000 samples taken in 10 ms with a 100 kHz sampling rate. In total we obtain an imaging rate of 15 Hz. For fibre 2 we use a stepping size of 1 um and otherwise the same parameters giving a sampling rate of 13 Hz due to the longer step size. During the 302 scans of 600 um length the positioning system measures its own slow drift to a total of 15 um. To have an equal number of measurements at each position we bin the data in 570 bins of 1 um for the statistical analysis, discarding the outer 15 um at each end. The transmission drops 2.1$\%$ during the scans. To compensate for this, the 302 scans were normalised to the mean of the 100 right-most points.  After this we use the 450 left-most points to determine the relative noise and find a standard deviation of $0.70 \text{ nm} \pm 0.05 \text{ nm}$. 

Scanning electron microscope (JEOL 7100 FEG) imaging of the tensioned ONF is made harder by charging effects as well as electron driven mechanical motion. These effects are alleviated by placing the ONF onto a silicon substrate, and imaging the fibre with low SEM current. The SEM pictures are taken with 30.000 times magnification and 10 kV acceleration voltage. We digitally analyse the SEM images by their brightness, counting pixels which are brighter than twice the average dark noise as fibre. The images have a 960 x 1080 pixels resolution with each pixel corresponding to a scale close to 3 nm x 3 nm. To compare with the scan data we average each image to one data point. The resulting width of the nanofibre is quite sensitive to the choice of threshold value, leading to a systematic radial uncertainty of $\pm 10$ nm in the waist region for a threshold value of 1.33 to 2.67 times the average dark noise. 
\begin{figure}[htb!]
\includegraphics[width=0.41\textwidth,keepaspectratio]{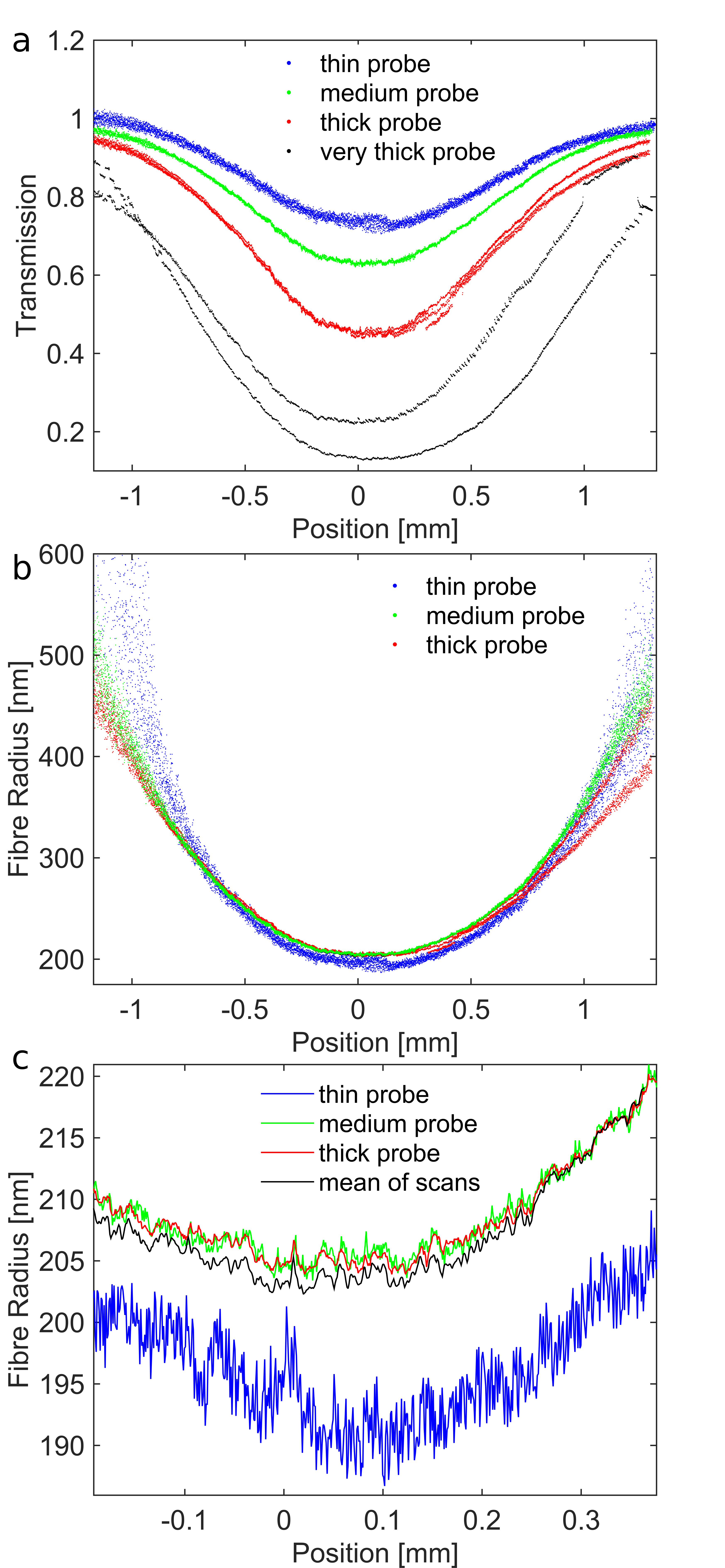}
\caption{Resolution and hysteresis as function of probe thickness. a) Transmission measurements for 4 probe thicknesses. The data of the medium and thick probes has been compensated for left-right hysteresis of 5um and 6um respectively (green, red). b) Resulting sample fibre radiuses with $\eta=53\% , 82\%$ and $121\%$ of the outside intensity scattered. c) Zoom-in on central region of b), plotting one scan taken with each probe thickness (thin, medium and large), as well as the mean from the scans in figure 4 b). The data points have been connected to aid the eye. 
}
\label{fig2sup}
\end{figure}

\section{Additional data}
The imaging performance with thicker probes is investigated in Supp. Figure \ref{fig2sup}. We use 4 different positions on the tapered probe fibre providing 4 different probe thicknesses. Figure \ref{fig2sup} a) shows that each probe thickness gives a different maximum scattering at the sample waist with the thicker probes scattering more light as expected. The thinnest probe (blue) gives a lower signal-to-noise ratio and we observe artefacts in 3 of the 4 scans with magnitude 0.02 in transmission. The green points correspond to the data presented in the main text in Figure 4. The thick probe (red points) shows a few artefacts; especially we see an undesired jumping near 0.4 mm and a systematic offset between repeated scans on the far right. The offset on the far right is expected to be temporary deformation caused by the drag. Despite these drawbacks the trace displays a very low noise. The thickest probe (black) completely deforms the fibre so even though it might have a very low noise it is not suitable for imaging the fibre and so it is only displayed in Figure  \ref{fig2sup} a).  In Figure \ref{fig2sup} b) we have modelled the radius with Eq 1 of the main text, using $\eta$ of $53\%, 82\%$ and $121\%$ for the thin, medium and thick probes respectively, each found by normalising at the single mode point. The scattering up to $\eta=121\% \textgreater 100\%$  can be explained by coupling between the inside and outside power in the probe region, thereby allowing more than the fraction of power in the evanescent field to be scattered. The data shows that the model still gives the same prediction for the radius in this regime. In Figure  \ref{fig2sup} c) we zoom in the on the waist region. Here it is clear that the mean of the 302 scans with the medium probe (black trace) and the single pass thick probe (red trace) are in very good agreement showing that our method can reach a resolution of the order of magnitude of 0.1 nm in a single scan.

\begin{figure}
\includegraphics[width=0.45\textwidth,keepaspectratio]{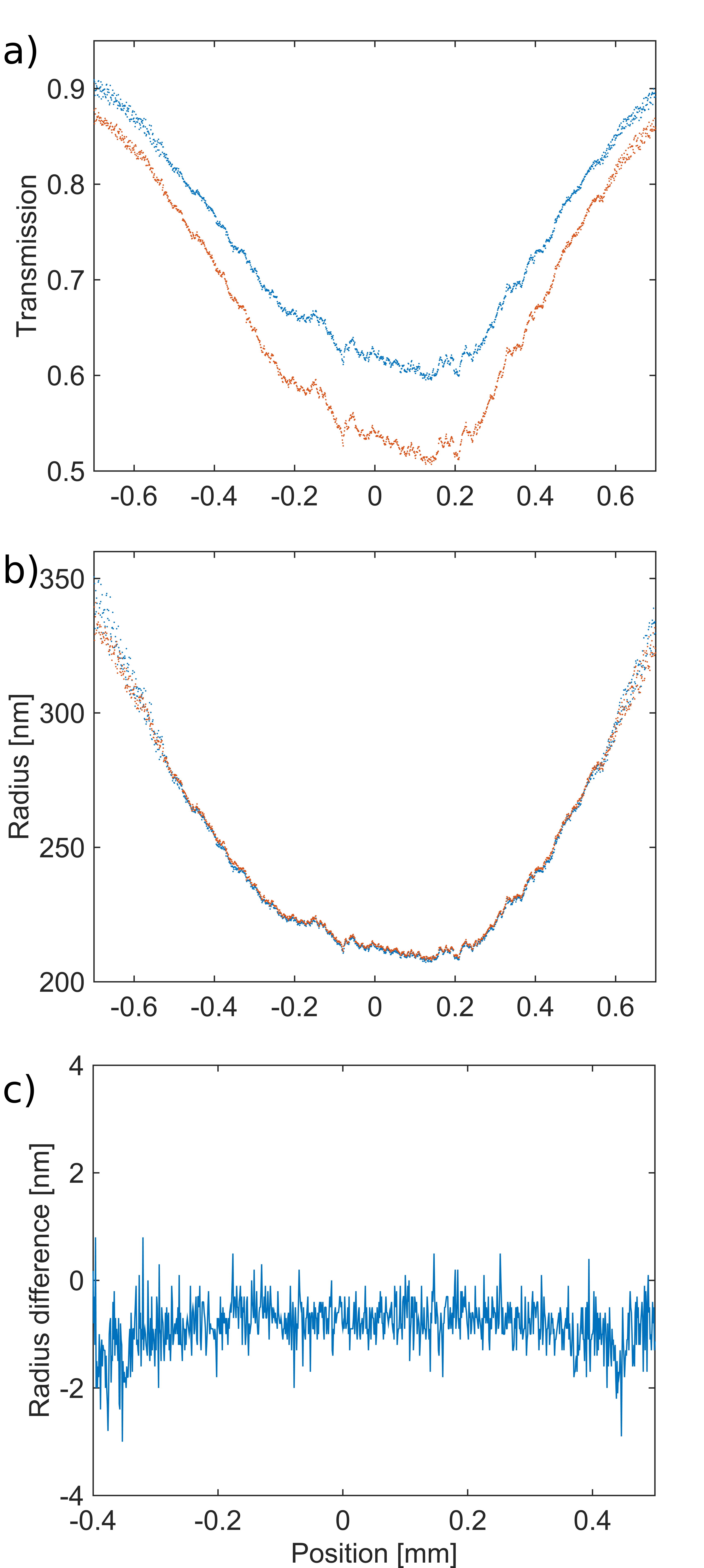}
\caption{Polarisation data. a) Transmission  measurement with the polarisation set to give maximum and minimum transmission on the waist for a single probe thickness. b) Radius calculated from the same data as a) with $\eta=92\%$ and $\eta=113\%$ respectively. c) The difference in radius between the traces from b).
}
\label{fig4sup}
\end{figure}

\begin{figure*}[htb]
\includegraphics[width=1\textwidth,keepaspectratio]{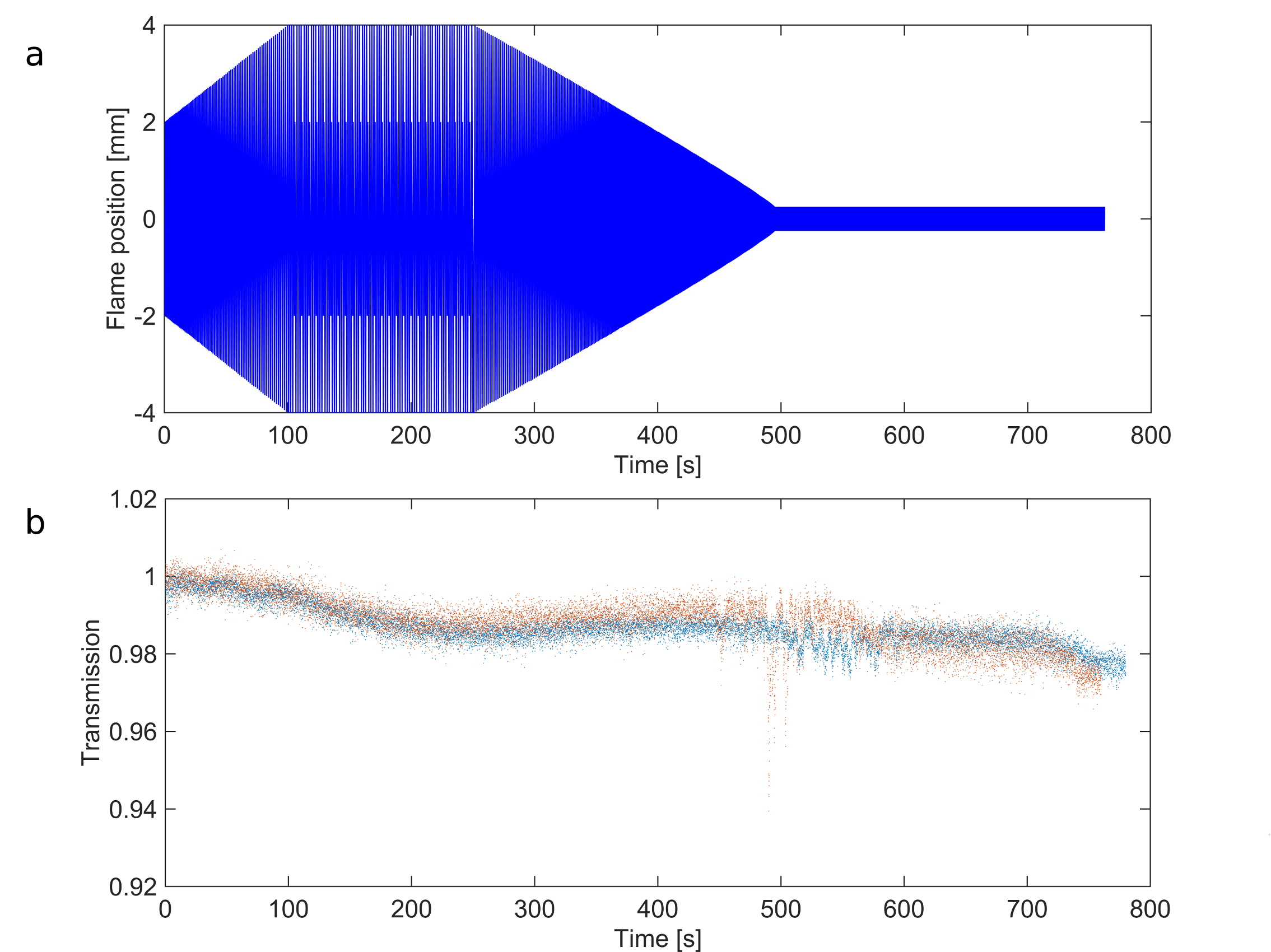}
\caption{Fabrication data. a) Pattern used to dither the flame along the y-axis while pulling the sample fibres. b) Time traces of the normalised transmission while pulling the sample fibres. Blue points are the first sample fibre and the red points are from the second sample fibre.
}
\label{fig3sup}
\end{figure*}
The polarisation stability of the method and the birefringence of the nanofibre is investigated in Figure \ref{fig4sup}. The input polarisation of the probe light is controlled with a 3 paddle fibre polarisation rotator. The probe is placed in the middle of the sample and the polarisation is set to either maximum or minimum transmission, resulting in a difference of 10\% in transmission. The sample is then imaged for the two settings of the polarisation, see Figure \ref{fig4sup}a. While changes in sample thickness give correlations in the two traces, birefringence would rotate the maximum and the minimum towards each other giving anti-correlations in their relative separation. To test the methods robustness to input polarisation we derive the radius with the result shown in Figure \ref{fig4sup}b. Normalization makes sure that the fibres have close to the same thickness in the single mode point.  Birefringence and drifts in input polarsation during the measurement would make the two traces drift apart. To magnify the drift we take the difference between the traces shown   \ref{fig4sup}c. The systematic offset comes from the individual calibrations of the single mode point. The maximum difference that the polarisation could cause is 35 nm whereas the observed separation is less than 2 nm. This shows that there is no significant build up of birefringence or drift in input polarization. On the one hand setting the probe to minimum and maximum transmission gives the greatest robustness to such drifts by suppressing the effect of polarisation to first order. On the other hand the sensitivity to birefringence can be maximized by setting the polarisation to two orthogonal polarisation states centered between the maximum and minimum.

The fibre puller setup is built around a hydrogen torch on a Newport M-ILS150-ccha stage burning 300 sccm of pure hydrogen (Alicat flowmeter). An optical fibre (Thorlabs 780hp) is stripped and cleaned with acetone. The fibre is held in place by two fibre clamps. The hydrogen torch is dithered back and forth along the y-axis using the pattern shown in Fig. \ref{fig3sup} a) while streching the fibre at constant speed. The distance between flame and fibre along the x axis is kept constant in the first 550 seconds, and then moved back 0.2 mm to minimize air turbulence in the final stage of tapering. 

While tapering the transmission is measured and recorded with a 20 Hz sampling rate. The power is stabilized as in the main experiment. In Figure \ref{fig3sup} b) traces for the two fibres presented in the main text are shown. The onset to multimode operation around 500 seconds is slightly offset and the traces are stopped at slightly different times. This is caused by minor changes in the setup associated with 3 months in between pulling the two fibres and the setup being moved. From the point of view of transmission the final results after the tapering are very similar with close to 98$\%$ transmission.

\lstset{language=Matlab,%
    breaklines=true,%
    morekeywords={matlab2tikz},
    keywordstyle=\color{blue},%
    morekeywords=[2]{1}, keywordstyle=[2]{\color{black}},
    identifierstyle=\color{black},%
    stringstyle=\color{mylilas},
    commentstyle=\color{mygreen},%
    showstringspaces=false,
    emph=[1]{for,end,break},emphstyle=[1]\color{red}, 
    columns=fullflexible,
    upquote=true,
    literate={-}{{-}}1
   {*}{{*}}1
}

\section{MATLAB Code}
The theory needed to model the fibre has been developed by \citeauthor{snyder2012optical,le2004field} and in great detail by \citeauthor{vetsch2010thesis}  which we follow here\cite{snyder2012optical,le2004field,vetsch2010thesis}. We find the propagation constant of the guided mode as function of fibre thickness and use this to obtain\cite{vetsch2010thesis} the fraction of power guided inside $\frac{P_{in}(r)}{P_{total}(r)}$ and outside $\frac{P_{out}(r)}{P_{total}(r)}$ of the fibre. Finding the propagation constant is a numerical task, so to efficiently make the conversion between transmission and inferred radius we use a lookup table leaving $\eta$ as the only fitting parameter. The MATLAB code below generates the table.
\begin{lstlisting}
clear all
close all

%% Parameters
%wavelength
lambda=780*10^-9; 
%wavenumber
k_in= 2*pi/lambda; 

%refractive indices
%glass (fibre)
n1= 1.46; 
% air
n2= 1; 
%fibre radii from 175 nm
%to 700 nm in 0.1 nm steps
fiber_radius=175*10^-9:4*10^-10:700*10^-9; 
%resolution of probagation constant of fibre,
resolution1=10000;
%propagation constants from n2*k_in to n1*k_in
beta=linspace(1.01*n2*k_in, ...
    n1*k_in*0.99,resolution1); 
%% 
for fff=1:length(fiber_radius)
%fibre radius
a=fiber_radius(fff); 

%% Propagation constant
%Designed for finding the fundamental mode
a_vs_beta=abs(besselj(0,sqrt(k_in^2*n1^2 ...
-beta.^2)'*a)...
./(besselj(1,sqrt(k_in^2*n1^2-beta.^2)'*a)...
.*(sqrt(k_in^2*n1^2-beta.^2)'*a))... 
-((n1^2+n2^2)/(2*n1^2)*...
(besselk(0,sqrt(beta.^2-k_in^2*n2^2)'*a)...
+besselk(2,sqrt(beta.^2-k_in^2*n2^2)'*a))...
./(2*(sqrt(beta.^2-k_in^2*n2^2)'*a)...
.*(besselk(1,sqrt(beta.^2-k_in^2*n2^2)'*a)))...
+1./(sqrt(k_in^2*n1^2-beta.^2)'*a).^2 ... 
...%-+ for finding other guided modes
-sqrt(((n1^2-n2^2)/(2*n1^2)*...
(besselk(0,sqrt(beta.^2-k_in^2*n2^2)'*a)...
+besselk(2,sqrt(beta.^2-k_in^2*n2^2)'*a))...
./(2*besselk(1,sqrt(beta.^2-k_in^2*n2^2)'*a)...
.*(sqrt(beta.^2-k_in^2*n2^2)'*a))).^2 ...
+(1*beta'/(n1*k_in).*...
(1./(sqrt(beta.^2 - k_in^2*n2^2)'*a).^2 ...
+1./(sqrt(k_in^2*n1^2-beta.^2)'*a).^2)).^2)));
%minimizing the result
% this is for single mode fibres case
if a<n1*0.40*lambda 
[~,dummy4]=min(a_vs_beta); 
dummy3=0;
%this is needed if there is a higher order mode
else 
[~,dummy3]=max(a_vs_beta); 
[~,dummy4]=min(a_vs_beta(dummy3:end));
end
beta11=beta(dummy4+dummy3);

%% Fraction of power in and out
h11=sqrt(k_in^2*n1^2-beta11^2);
q11=sqrt(beta11^2-k_in^2*n2^2);
s11=(1/(h11*a)^2+1/(q11*a)^2)...
/((besselj(0,h11*a)-1/(h11*a)...
*besselj(1,h11*a))/(h11*a*besselj(1,h11*a))...
-0.5*(besselk(0,q11*a)+besselk(2,q11*a))...
/(q11*a*besselk(1,q11*a)));

% fraction P_in/P_total=D_in 
%and P_out/P_total=D_out
D_in(fff)=(1-s11)*(1+(1-s11)*beta11^2/h11^2)...
*(besselj(0,h11*a)^2+besselj(1,h11*a)^2)...
+(1+s11)*(1+(1+s11)*beta11^2/h11^2)...
*(besselj(2,h11*a)^2-besselj(1,h11*a)...
*besselj(3,h11*a));
D_out(fff)=besselj(1,h11*a)^2/...
besselk(1,q11*a)^2*((1-s11)...
*(1-(1-s11)*beta11^2/q11^2)...
*(besselk(0,q11*a)^2-besselk(1,q11*a)^2)...
+(1+s11)*(1-(1+s11)*beta11^2/q11^2)...
*(besselk(2,q11*a)^2-besselk(1,q11*a)...
*besselk(3,q11*a)));
end
 %gives the theory curve of inset Fig. 3a
 %in the main text, single mode point at
 %transmission 0.81
eta=1-(1-0.81)/0.19; 
figure
plot(fiber_radius,(D_in+eta*D_out)./(D_in+D_out))
xlabel('Fibre radius [m]')
ylabel('Normalised transmission')
\end{lstlisting}

\bibliography{nearfield}